\documentclass{article}
\usepackage{graphicx}
\usepackage{amsmath}

\input{tcilatex}

\begin{document}

\title{The Title of a Standard LaTeX Article}
\author{A. U. Thor \\
The University of Stewart Island}
\maketitle

\begin{abstract}
We study the effects of warm water on the local penguin population. The
major finding is that it is extremely difficult to induce penguins to drink
warm water. The success factor is approximately $-e^{-i\pi }-1$.
\end{abstract}

\section{Figures caption:}

Fig 1: The dipolar autocorrelation function for different coupling parameter:

(---) weak coupling $\Gamma =0.125,$

(- - -) mean coupling  $\Gamma =1.25,$and 

\ (.....) strong coupling $\Gamma =2.5.$

Fig 2: Lyman alpha line (---) PPP code, (....) our calculation, T$_{e}=5\ast
10^{3}%
{{}^\circ}%
K,$ N$_{e}=10^{20}cm^{-3},$ $\Gamma =0.125$

Fig 3: Lyman alpha line (---) PPP code, (....) our calculation, T$_{e}=10^{4}%
{{}^\circ}%
K,$ N$_{e}=10^{20}cm^{-3},$ $\Gamma =1.25$

Fig 4: Lyman alpha line (---) PPP code, (....) our calculation, T$_{e}=10^{5}%
{{}^\circ}%
K,$ N$_{e}=10^{20}cm^{-3},$ $\Gamma =2.5$

\end{document}